\documentclass{nature-hack}

\usepackage{amsmath} %for split eq.
\usepackage{graphicx}
\usepackage{setspace} %for fig caption line space adjustment
\usepackage{csquotes}
\usepackage{amssymb}
\usepackage{xfrac}

\usepackage{tikz}

\usepackage{floatpag}
\floatpagestyle{empty}

\usepackage{nicefrac}

\usepackage[colorlinks,citecolor=black,linkcolor=black,urlcolor=blue,bookmarks=false,hypertexnames=true]{hyperref} 

\newcommand{\env}[1]{{#1}^{\:\prime}}

\newcommand{\Expect}{{\rm I\kern-.2em E}}

\title{Mutation Effect Generalizability under Selection-Drift} 

\author{{Andre F. Ribeiro$^{1}$ (ribeiro@alum.mit.edu)}}

\begin{document}

\maketitle

%\bigskip\bigskip\bigskip\bigskip\bigskip

\begin{affiliations}
\item{University of Sao Paulo, Sao Carlos, SP, 13560-970,  Brazil}
%\item{Department of Genetics, Evolution and Environment\\
%Centre for Life's Origins and Evolution\\
%University College London\\
%Gower St, Bloomsbury, London WC1E 6BT, United Kingdom}
\end{affiliations}

\begin{abstract}
%, and the consequent diversity,
%[diversity of] 
While Neutral Theory famously describes the number of discrete genetic differences in populations, we consider the number of genetic backgrounds under which such differences are observed - setting limits to the generalizability of their effects. This allow us to determine which population structures and diversity rates have maximal effect generalization across (1) environmental and (2) genetic variation, and to demonstrate that they correspond asymptotically to those of populations under (1) natural selection and (2) drift. At the same time, these results suggest distinct limits to the predictability of fitness and evolution across evolutionary regimes. We employ both broad time, large-scale genome sequencing datasets (including whole-genome autocorrelation calculations), and fine time-scale barcoding experiments.

%These results finally suggest strict differences to the generalizability of mutation effects, and the predictability of evolution, under these evolutionary regimes.

%Distinct population structures offer their populations adaptations with distinct robustness (i.e.,  effect generalizability across members). We demonstrate that populations under (1) drift and (2) natural selection have population structures and genetic diversity rates that correspond to the maximal-generalization structures and rates of, respectively, (1) stationary (constant) and (2) non-stationary (changing) environments. We employ both broad time, large-scale genome sequencing datasets (including whole-genome autocorrelation calculations), and several fine time-scale barcoding experiments.

%new limits to the prediction of fitness and evolution 

% at unprecedented scales
\end{abstract}

\section{Introduction}
%,Kryazhimskiy:2014tu
%across populations
% (as well as to the mediating role of population structures)
In complex genomes and environments, the effect of a mutation observed for a population member is rarely the same as that observed for another, due to gene-gene (epistasis) and gene-environment interactions. This could make the selection of beneficial mutations, and natural selection generally, scale poorly under environmental and genomic changes. Increasing evidence points to the shaping influence of mutation robustness (i.e., the external validity of mutation effects across population members) on the evolutionary process\cite{Zheng:2020uy,Johnson:2019ts,Chaturvedi:2021tr,Hernando-Amado:2022ut,Stewart:2022ul,Park:2022ul,Scheuerl:2020ui,Brennan:2019vp}. Experiments relating selection strength and robustness have offered qualitative-only observations of their relationship, making it difficult to generate new testable predictions, and articulate the relationship between robustness and traditional evolutionary theory. Robustness is typically quantified in simple ways in most studies\cite{Felix:2015we,Fares:2015ts} (e.g., the number of members with overlapping phenotypes in a population), but is related to quickly developing issues of generalization in Machine Learning and Statistics. Theoretically, the relationship between the evolutionary process and robustness\cite{Draghi:2010aa,Rao:2022vb,LaBar:2017we,Wilke:2001vb,Krakauer:2002wo,Nimwegen:1999ut}, or, the process and arbitrary population-structures\cite{BenjaminAllen2017Edoa,Pavlogiannis:2018aa,Lieberman:2005aa} have remained highly idealized, often accompanied of only simulated or, more commonly, no evidence. Here, we review the theory behind robustness, its relation to adaptation under pleiotropy, and reconsider multiple recent data in light of the resulting model. 
%and popular 

\begin{figure}
\centering
\includegraphics[width=1\linewidth]{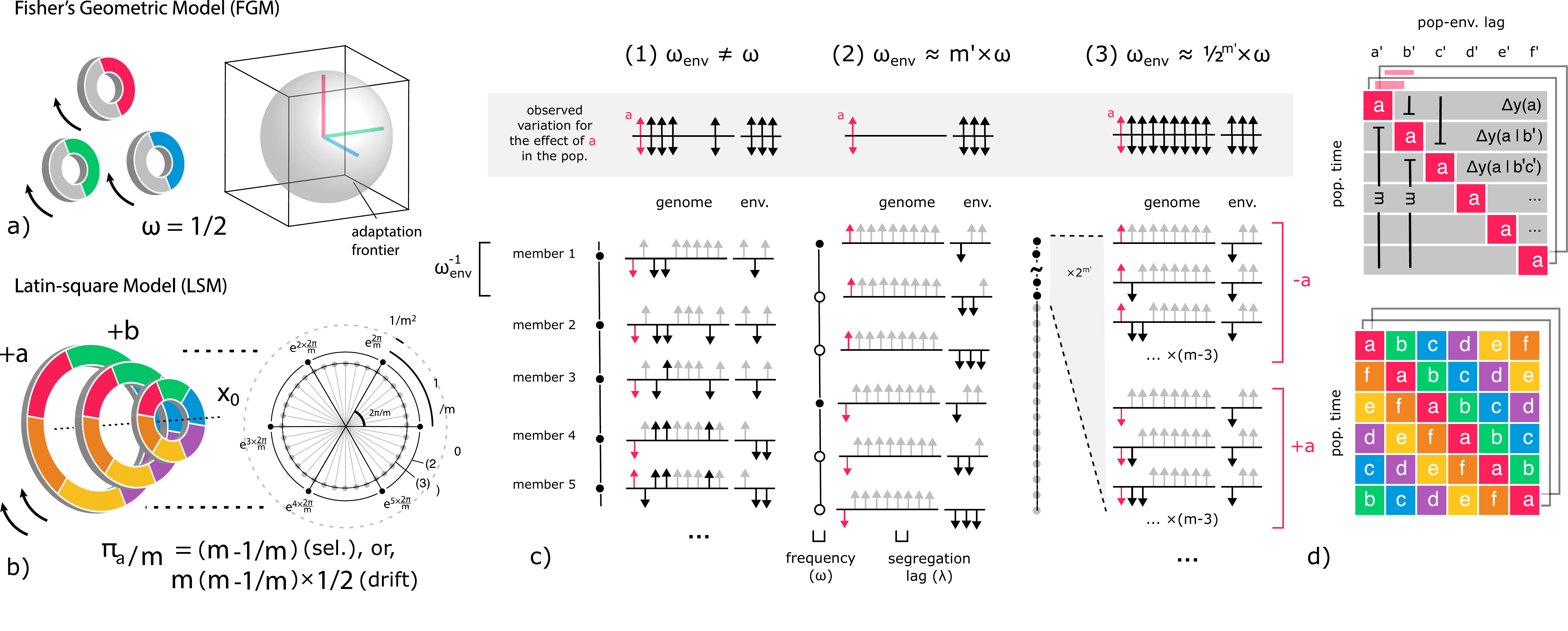}\\
{\small \refstepcounter{figure}\label{fig-model}\setstretch{0.5}\sffamily\noindent\textbf{Figure \arabic{figure}}\hspace{1em}\textbf{Latin-Square Model (LSM).} \textbf{(a)} Fisher's Geometric Model (FGM) and microscope analogy, \textbf{(b)} LSM as a mechanism to generalize mutation effects across populations, \textbf{(c)} population-environment systems $(1{-}3)$ with rates of change of $\omega_{env}$ for the environment and $\omega$ for populations, horizontal lines illustrates genomic and enviromental variants across population members (rows), arrows show whether they take the same value (up, $-1$) or alternate (down, $+1$) to a reference $x_0$, the gray box shows the range of variation under which the effect of a SNP $a$ is observed, systems' rates of genetic background generation lead to differences in mutation robustness across scenarios, \textbf{(d)} time-extended matrix representations for systems $(2)$ (top) and $(3)$ (bottom).}
\end{figure}
% (pop.-env. lag vs. time)
%, and its whole genome frequency-based representation (right)
%$\lambda$ is the inter-SNP lag across genomes, 

%horizontal lines show genomic and enviromental variants across members of an example population (rows), arrows show whether they take the same value 

%the microscope's
%which generates the full range of knob variations by phasing their rotations.
We can introduce the problem using Fisher's famous allegory for multi-dimensional adaptation with pleiotropy\cite{Fisher:1930wy,Tenaillon:2014ux,Orr:2005wt}.  An individual organism's fitness at an instant is a function of its traits (e.g., body size, beak length), like a microscope's of its knobs' positions. The process of random, and time-extended, adaptation is then analogous to one where we randomly change knob positions, until a sharp image comes through, Fig.\ref{fig-model}(a). Fitness gains and losses brought by any mutation are, however, contingent on a large number of factors (environmental, developmental, regulatory, etc.) Each observation of fitness is thus valid only in its very instantaneous set of conditions. Since gains from a knob depend on the position of all other knobs, the only sure-way to assure no false-positive adaptations is to try each knob position, in each variation of all other knobs. This is illustrated by the wheel in Fig.\ref{fig-model}(b), which generates all possible knob variations by phasing their rotations. We call this alternative to Fisher's Geometrical Model (FGM) the Latin-Square Model (LSM). 

%haploid populations,
The FGM is a full randomization approach, where we randomly switch knobs, and approach the instrument non-methodically. The LSM is closer to how a microscope is actually used, where we change a given knob, while systematically fixing all other knobs' values, in a 'structured randomization' approach that is akin to simultaneous experimentation. While the FGM makes many assumptions (strong stabilizing selection, equal effect mutations, etc.) the one that washes away issues of robustness is that there are no mutational correlations among traits\cite{Orr:2005wt,Fisher:1930wy}. In more realistic conditions, moving a knob would move many other knobs, in unknown ways. The microscope's design, like the experimental, is an apt abstraction because it breaks down all possible variations available to its operator in dimensions whose effects remain unconfounded throughout adaptation. Exactly because of this assumption of independence, Fisher was able to model adaptation as trajectories in a Euclidean $m$-dimensional space, finally showing that adaptation largely takes place in a sphere around the optimal, Fig.\ref{fig-model}(a). In both models, FGM and LSM, mutation effects are commutable. In Fisher's model, this requires a statistical independence assumption, while, in the LSM, commutation is afforded by population structure - taken as part of the adaptation process itself. Such questions are fundamental, as they cut to the core of what selection and adaptation are about: populations' ability to quickly identify the effects of previously unseen, or 'untested', mutations. As the main source of genomic changes, mutations also play a key role in explaining how populations use genomes to collectively represent (and react to) their environments. This vital population-environment connection has been increasingly de-emphasized with the availability of packaged omnic data, but is formulated explicitly in the LSM. 

%The LSM, further introduced below, has five key advantages, which make it uniquely suited to empirical research,

%\begin{itemize}
%\item[-] \textbf{Non-parametric.} The LSM is a direct implication, and not a theoretical model, of the population-environment combinatorial structure. It requires and makes no complex parametric assumptions.

%  \item[-] \textbf{Effects from the ground-up}. Theoretical models often have hypotheticals (fitness landscapes, phenotypic multi-dimensional spaces, effective populations, phenotype-genotype graphs, etc.) that make them difficult to validate. The LSM is formulated entirely from (correlated, confounded, etc.) observed effects of mutations on phenotypes, which are readily observed in experimental datasets, and estimated in observational.  
  
 % \item[-] \textbf{Full-genome scale}. Models of adaptation often consider mutations at a single site, gene or small genome. Mutation rates are, however, a population characteristic, and periodic or multi-scale patterns, like the ones below, are biased, phased, or disappear in genomic segments. Using large-scale computation, we consider combinatorial patterns across thousands whole-genomes in populations. 

 % \item[-] \textbf{Representational role of single genomes.}  

% \item[-] \textbf{Theoretically consistent}. The LSM is consistent with mainstream models; being, in fact, a statistical re-interpretation of Moran and Wright-Fisher models (and the FGM)\cite{sm-anonymous}, while adding to these models sophisticated statistical tools to articulate and quantify populations' mutation robustness.
%\end{itemize} 

\subsection{Populations under external and stationary change.}

%for each variant
%rate $\omega_{pop}<\omega_{env}$
%When population and environment are in-phase, (1), every change is instantaneously selected across all populations.
%the population has a member with $a$ every $1/m$ (of an environmental cycle). In this case
%Choosing a varying lag, instead, generates effect observations under alternative conditions.
%, and environment-population systems that are reciprocally periodic
Consider a population in an environment that is changing at a rate of $\omega_{env}$. A critical question for that population is how to choose, in response, its own rate of change $\omega$. Together, the population and environment make up a reciprocal system, where the environment has $\env{m}$ variations, $\{\env{a},\env{b},\env{c},...,[\env{m}]\}$, that can affect population fitness, $y \in \mathbb{R}$ (where $[m]$ indicates the $m$-th variant). Throughout this article, we use Latin letters, $\{a,b,c,...\}$, to refer to population genetic variations (Single Nucleotide Polymorphisms, SNPs) and primed letters, $\{\env{a},\env{b},\env{c},...\}$, to environmental. We also write ${+}a$,${-}a$ and ${\pm}a$ for the presence, absence and polymorphism of $a$ in populations. Consider then how populations' chosen rates, $\omega$, affect their ability to adapt, in particular, in respect to the number and types of effects they observe. %as result of alternative $\omega$ choices. 

Fig.\ref{fig-model}(c) illustrates three scenarios

\begin{equation}\label{eq-priorrates}
(1) \;\; \omega \neq \omega_{env}, \qquad
(2) \;\; \env{m} \times \omega =  \omega_{env}, \qquad
(3) \;\; \nicefrac{1}{2^{m'}} \times \omega =  \omega_{env}.
\end{equation}

%equal proportion (abscence/presence)
%Fitness-relevant 
 Genetic variation is illustrated as ${-}1$ or ${+}1$ arrows over individual genomes in a population with $m$ segregating sites (left horizontal lines, values for a reference genome $x_0$ taken arbitrarily as $-1$) and $n$ members (rows). Environmental variation is shown similarly in the line to the right. Consider the problem of estimating the effect (on fitness $y$) of a SNP $a$. With rates $(1)$ these effects are doubly confounded. Most effect observations $\Delta y(a)$ made in such populations reflect uncontrolled variation in the  environment, or concurrent genetic variation. For example, the observed effect on fitness precipitated by population member $1$ in Fig.\ref{fig-model}(c) (first row and column) is the difference $\Delta \hat{y}(\,a\, |\, x_0,{\text{+}}c,{\text{+}}d,{\text{+}}\env{b} ) =  y(\, x_0\, |\, {\text{-}}a, {\text{-}}b,{\text{-}}c,{\text{-}}d, ..., {\text{-}}\env{b}, {\text{-}}\env{c}, {\text{-}}\env{d}, ...\,) - y(\, x_0\, |\, {\text{+}}a, \allowbreak{\text{-}}b,\allowbreak{\text{+}}c,\allowbreak{\text{+}}d,\allowbreak ..., \allowbreak{\text{+}}\env{b},\allowbreak {\text{-}}\env{c},\allowbreak{\text{-}}\env{d},\allowbreak ...\,)$. This effect observation reflects not only $a$'s effect, but also the extraneous variation of $(c,d,\env{b})$. Using such observations for the selection of $a$ could promote deleterious adaptions, and the increasing accumulation of deleterious epistatic effects. 

%(similarly to all others)

%value of  = y(x_0 | +a, +b,+c,+d, ... [m], b', c',d', ... [m]) - y(x_0 | -a, +b,+c,+d, ... [m'], -b', +c,+d, ... [m])$
%[At each enviromental instant, we observe fitness $y$ under both possible conditions, which allow us to estimate their effects and indepedent value.]
%[, as effects are observed over single enviromental variations] [and the slower enviroment\footnote{notice that because the environment is by definition slower, its effects are individual and not confounded.}]
%[, as they are valid for specific (a,b,c,d) values only]
In contrast, rates $(2,3)$ lead to population-environment systems where effects are not confounded (and to populations that change at, respectively, slower or faster frequencies than their environments). The gray top-panel in Fig.\ref{fig-model}(c) depicts all variation under which the effect of SNP $a$ (red) is observed in the previous population, from the individual-level observations in the lower rows. With a frequency of $\env{m}\times \omega_{env}$, we observe the effect $\Delta y(\,a)$ under every possible environmental variant, $\Delta \hat{y}(\,a \,| \, \allowbreak{\pm}\env{b},\allowbreak{\pm}\env{c},\allowbreak{\pm}\env{d}, ...\,)$, Fig.\ref{fig-model}(c, middle).  With such per-SNP frequency, the effect of $a$ is appropriately separated from the effect of other SNPs, due to a slow rate of change. Effects in $(2)$ are observed, however, in only one genetic background, $\Delta \hat{y}(\,a \,| \, {\text{-}}{b},{\text{-}}{c},{\text{-}}{d},...\,)$, repeatedly (across all times and populations). Any changes in factors $({b},{c},{d},...)$, are likely to invalidate these effect observations and, consequently, populations' ability to choose what individual adaptations to promote. We thus say that in $(2)$ effects are unconfounded, but not generalizeable across the population. In systems $(3)$, in contrast, each effect $\Delta y(a)$ is observed under full genomic and environmental variation, $\Delta \hat{y}(\,a\, |\, {\pm}b, {\pm}c, {\pm}d,...\,{\pm}\env{b},{\pm}\env{c},{\pm}\env{d}, ...\, )$, Fig.\ref{fig-model}(c, right). Any frequencies different from $(2)$ will lead to environmental changes unequally represented in populations (and be 'unbalanced' in the sense of Experimental Designs\cite{Montgomery:2001aa,ribeiro-ev}), and any frequencies different from $(3)$ will lead to genetic changes that not fully generalize across populations. Given the widespread prevalence of epistatic effects among SNPs\cite{Breen:2012tp,Park:2022ul,Kryazhimskiy:2014tu}, the generalizability of their effects should be a key factor for adaptation.  

%The expected value of effects correspond to the effect with maximum generalizability.

%Confoudeness is related to extraneous variation and EV to incomplete.

%In both cases, effects are generalized over enviroment conditions, but only in $(3)$ effects have high ACC [generalize across population].

%, or discrete genetic differences in a population\cite{Watterson:1975aa,Tajima:1983aa,Kimura:1983tf,Nei:1979uk,ribeiro-ev}
Define a quantity $\nicefrac{\partial \pi_{a}}{ \partial m}$ counting the number of effect observations per SNP across populations. The asymptotic limit for this quantity follow directly from the previous rates, Eq.(\ref{eq-priorrates}),
%Crucially, the previous population-environment systems, $(2,3)$, lead to distinct rates for populations.

\begin{equation}\label{eq-posrates}
%(1) \;
%\begin{cases}
%\frac{\partial \pi_{a}}{\partial t} = 1,\\
%\end{cases}
(2) \,
\begin{cases}
 \; \frac{\partial \pi_{a}}{\partial m} = \Big( 1 - \frac{1}{\env{m}}\Big)^{m'} \xrightarrow{} \frac{1}{e} = 0.36...,
\end{cases}
(3) \,
\begin{cases}
\; \frac{\partial \pi_{a}}{\partial m} = \frac{1}{2} \Big( m - \frac{1}{m}\Big)^{m'} \xrightarrow{} \frac{\phi}{2} = 0.81...,
\end{cases}
\end{equation}

%time extended

%The constant lag in system (1) enumerate effect observations in 1 bacground per variant.
%A system with periodic lag $m$ enumerate observations in $m$ distinct backgrounds.
%Any other lag-frequency combination would lead to biased fitness and effect observations across populations. 
%By considering rates per genome size $m$, we ignore reproduction rates in populations. 
%\footnote{$m \times (1-\nicefrac{1}{m})=(m-1)$.}
%, for SNP $a$,
%, $n> m$
%sampling

which describe the frequency under which effects of a particular SNP are observed every $m$-time (i.e., per-genome), and $\phi$ is the golden ratio ($\env{m}$ increasing, $m$ constant). The rate $(2)$ can be thought as: for any new segregating site, make only $1$ effect observation for each $m$ variant, observing the same effect across distinct environmental conditions, Fig.\ref{fig-model}(c, middle). The per-$m$ rate is $m \times (1 - \nicefrac{1}{\env{m}}) = (m-\nicefrac{1}{m})$ in this case, Eq.(\ref{eq-priorrates}). The rate for $(3)$ can be thought as: for each new site, also observe effects for all other $m-1$ sites, and $(m - \nicefrac{1}{m})$ positions every $m$ time\footnote{the rate $(3)$ is divided by $2$ to account for observations with and without $a$, as required to calculate the effect $\Expect [\Delta y(a) ]$ in populations, and can be calculated in the two cases at rates $(2)$ $2\times \omega_{env}$  and $(3)$ $\omega_{env}$, Fig.\ref{fig-model}(c).}. The per-$m$ rate for a fixed $a$ value is $m \times (m - \nicefrac{1}{m})$ in this case (i.e., $m$ times the previous). These two rates express opposite conditions for effect observations, where they are observed under none or all possible genetic backgrounds. With a fixed number of segregating sites, $m$, these two rates lead, in turn, to known asymptotic expressions for Euler's number and Fibonacci series rates\footnote{see, for example, \cite{boyer1968history}(Chapter 20) and \cite{Gazale:1999aa} for simple and illustrated discussions.}, Eq.(\ref{eq-posrates}). Alternative proofs are discussed in \cite{ribeiro-ev} (\textit{Sect. 6 Sample Power}), as well as relationships to random sampling and trees. 

%Unsurprisingly, the second rate for a fixed $a$ is simply a $m$ times repetition of the first\footnote{from Eq.(\ref{eq-priorrates}), $(2)$ $m\times (1-\nicefrac{1}{\env{m}})=m\times (1-\nicefrac{1}{m^2})=m -\nicefrac{1}{m}$, and, $(3)$ $m\times (m - \nicefrac{1}{m})$.}.

%***
% In this article, our focus is in demonstrating that rates $(2{,}3)$ are associated with populations under, resp., selection and drift, and the consequences of these alternative rates and structures for mutation effect generalizability across populations. 

%and periodic

%These are simple combinatorial consequences of the population-enviroment but they will have clear and long-term consequences for the generalizability of effects in populations. 

%longitudinal
Population-environment systems with rates $(2,3)$ in Eq.(\ref{eq-priorrates},\ref{eq-posrates}) implement the same process, but display different segregation and linkage disequilibrium patterns. We will study multiple populations with common size and time of evolution, but either under drift or selection, and demonstrate that their background diversity rates correspond to, respectively, $(2)$ and $(3)$. If the consequent number of effect background observations is a factor on mutation effect generalizability, then we should also observe distinct generalizability patterns in these two evolutionary regimes, across all studied populations. This relationship has been demonstrated in purely statistical grounds\cite{ribeiro-ev}, and is reconsidered bellow for both long and fine time-scale genetic data.

\subsection{Populations under external and non-stationary change.}

%A $m$-square's diagonal enumerates all conditions for one variant (stationary), the full square for $m$ variants (stationary), and all $m$-squares for $m$ variants in non-stationary environments

The definition of mutation robustness implied by the previous discussion is that of robustness as effect invariance. Effect invariance means that whether we make a change $a$ to a population $x_i$ or another $x_j$, the observed changes in fitness will be identical in both cases, $y(x_i+\{a\}) - y(x_i) = y(x_j+\{a\}) - y(x_j) = \Delta y(a)$, for any $x_i,x_j \subseteq X$ (where $X$ is a set containing all fitness-relevant genetic and enviromental factors). Any such measure requires, however, the enumeration of all possible background conditions, $x_i$, to add $a$. Systems with rates $(2,3)$ enumerates all conditions in stationary environments - i.e., with a constant $\omega_{env}$ in Eq.(\ref{eq-priorrates}). Repetition of the same cycles, one for each non-cyclic permutation of environmental variants\cite{ribeiro-ev}, are necessary in non-stationary environments. Systems $(2,3)$ correspond, therefore, to the only systems where effect invariance for one, $(2)$, or many, $(3)$, mutations can be evaluated and optimized. In them, invariance can be quantified simply by effect observation variance after full background variation, $\textrm{VAR}^{-1}[\,\Delta y(a \;|\; \Pi(X-\{a\}) )\,]$ (where $\Pi(X)$ is the set of all permutations of the elements in $X$\cite{ribeiro-ev}). That is, a system is said to be effect-invariant if it is able to \textbf{sustain small effect variance under full external variation}. The rates in Eq.(\ref{eq-priorrates},\ref{eq-posrates}) can thus be seen as generating population structures that maximize invariance, and robustness, under stationary conditions - and an equilibrium position for populations in non-stationary environments. 

%Squares can thus be seen as population structures maximizing invariance, and robustness, under stationary conditions - and an equilibrium position for populations under external changes. 

%LSM matrices are circulant matrices\cite{sm-anonymous,Pollock:2009vs}, which have simple periodic representations, Fig.\ref{fig-model}(b, right), with frequencies $\sfrac{1}{m}$ (outer circle) for system (2) and $\sfrac{1}{m^2}$ (inner circle) for (3). The representation leads to equivalent statistical and algebraic interpretations for mutation robustness. Statistically, a system is effect-invariant if it is able to \textbf{sustain small effect variance under full external variation}. This can be indicated by a null Generalized Variance (GV)\cite{sm-anonymous}, the determinant of the population's variance-covariance matrix, but only in the previous systems (i.e, with their rates and stationarity assumptions).  

%Algebraically, the square matrix is invariant under changes of basis. The previous definitions lead thus to a coordinate-free definition of effects, as effect observations that remain invariant under all possible changes of basis of the type $x_0-{a}$ to $x_0+{a}$. Both perspectives lead to factorwise linear-systems asymptotically (i.e., towards population effect homogeneity), while allowing for non-linearity outside equilibrium. This is a common pattern in Biological and Economic population-environment systems. 

\section{Results}

\begin{figure}
\centering
\includegraphics[width=1\linewidth]{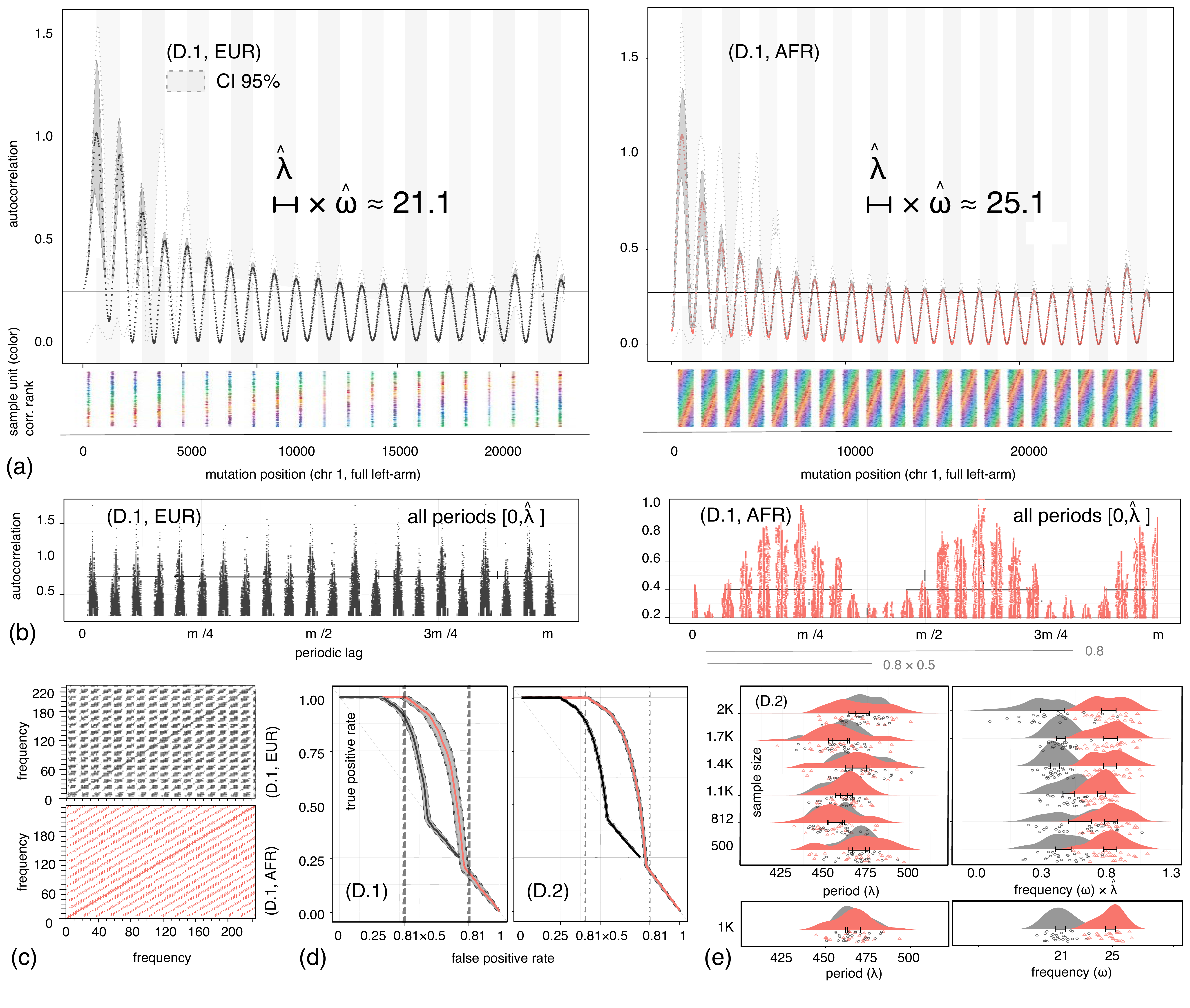}\\
{\small \refstepcounter{figure}\label{fig-1}\setstretch{0.5}\sffamily\noindent\textbf{Figure \arabic{figure}}\hspace{1em}\textbf{UK Biobank and 1K Genomes (broad time-scale).} \textbf{(a)} Autocorrelation $\rho[seg(i)]$ over left-arm of human chromosome $1$ across European (EUR, $n=503$, left) and African (AFR, $n=661$, right) populations of the 1K Genomes project\cite{Auton:2015vv} (D.1), gray-white bars show empirical number of cycles per genome, $\hat{\omega}$, and their lag period, $\hat{\lambda}$, autocorrelation confidence intervals (CIs) are shown as dark gray ribbons, \textbf{(lower strips)} population (color) correlation rankings across all sites shows periodic correlation, \textbf{(b)} genomic periodicities for EUR and AFR populations according to cumulative autocorrelations $\rho[seg(i)-seg(i+l)]$ over all integral lags $0< l \leq \hat{\lambda}^2$, \textbf{(c)} graphical test for periodic correlation, \textbf{(c)} SNP survival Empirical ROC curves follow the rates in Eq.(\ref{eq-posrates}), \textbf{(d)} UK Biobank\cite{Halldorsson:2022ux} (D.2)'s UK and African born populations (bootstrap with sample sizes of $[500,2000]$), \textbf{(bottom-row)} have distinct $\omega$ but common $\lambda$, and,  \textbf{(top-rows)} distinct periodic components following Eq.(\ref{eq-posrates}). }
\end{figure}

%over both $\pi$ distinct differences, $a,b,c...$, and genetic backgrounds, $x_0$
 The number of genetic differences maintained in populations, $\pi$, is a key quantity in \textbf{Neutral Theory} (NT)\cite{Watterson:1975aa,Kimura:1983tf,Tajima:1983aa}. Under NT, for a constant size (diploid) population at equilibrium with $m$ segregating sites\cite{McCandlish:2014us}, $\Expect [\pi]=\Expect\left[\sfrac{m}{\sum_{i=1}^{n-1} \frac{1}{i}}\right]=4N\mu$. Fisher's Geometric Model (FGM), in turn, has the historical significance of being the first mainstream model to justify the binomial rates observed for genetic differences in populations under no selection pressure, later leading to NT. As expected by NT, rate $(3)$ in Eq.(\ref{eq-priorrates}) leads to binomial distributions for the number of genetic differences, and selective sweeps with exponential rates, $(2)$, to overdispersed distributions. Unlike these well-know measures of nucleotide and haplotype diversity, we considered the number of possible backgrounds observed for each fixed difference, and interpreted them as counts of SNP effect observations. The consequent patterns in Eq.(\ref{eq-posrates}) are patterns over background counts. This characterizes population with a pair $(\pi,\nicefrac{\partial \pi_{a}}{ \partial m})$ of variables, and  leads to counts of partial permutations in populations (\textit{Sect. Combinatorial Enumeration}), and not just their differences. The relationship to \textbf{Wright-Fisher} and Moran models of drift are also discussed in the Supporting Material\cite{sm-anonymous}. In this section, we study how background counts change empirically across evolutionary regimes, and their statistical consequences to populations. Understanding how to predict mutation effects (fitness gains and losses) from environment and genetic conditions has, not only theoretic, but also great practical consequences\cite{Wang:2018uf,Nosil:2020tf,Nosil:2018vb}.

%The difference of rate $(3)$ in Eq.(\ref{eq-posrates}) to NT/FGM rates is the $\phi$ term, which describe rates necessary for full background variation of SNPs.  
 
 %[These results thus suggest statistical and strategic contexts for NT's distributions and rates, which is, largely, a descriptive theory.]
 
 %the effect observation requirement of
 %This suggests a justification for NT distributions and rates, which are largely descrptive.

 %This celebrated relationship has been re-formulated across theories\cite{McCandlish:2014us}, but can also be interpreted in combinatorial grounds, as  $\sum_{i=1}^{n-1} \frac{1}{i}$ is the expected number of fixations (overlaps) among random permutations. The case of large $\pi$ (compared to $S$) is associated with random diversity rates and the Binomial distribution described by drift and NT\cite{Tajima:1983aa,sm-anonymous}. Small $\pi$ is associated with exponential rates and overdispersed distributions that often characterize populations under selection. %, which have been more difficult to model. 

\subsection{Whole-genome Autocorrelation.}
%,Tajima:1983aa
%In fact, the 3-way relation between population diversity, growth and fitness leads fundamentally to periodic functions\cite{Ribeiro:2021aa}.
 
%these quantities and
We first consider evidence for the LSM using large-scale sequencing datasets, then experimental data. Models of adaptation often consider mutations at a single site, gene or small genome. Mutation rates are, however, a population characteristic, and periodic or multi-scale patterns, like the ones below, are biased, phased, or disappear in genomic segments. Using large-scale computation, we consider combinatorial patterns across thousands of whole-genomes in populations. Fourier and frequency-based representations have been central to many key scientific discoveries (e.g., the DNA double-helix and quantum double-slit experiments). Although the relevance of frequency-based representations to omnic data has long been hypothesized\cite{Weinberger:1991we,Poelwijk:2016vj}, it has led to few empirical results.  Genome autocorrelation is a (circular) convolution of the full genome with itself, and indicates across-genome patterns of pairwise overlap and differences - making them useful to demonstrate the previous combinatorial quantities and limits. The conditions in Eq.(\ref{eq-posrates}) imply that drift-selection transitions change the frequencies of effect observations, $\omega$, across populations' genomes, while keeping the separation between segregating sites, $\lambda$, constant. These two quantities are illustrated in Fig.\ref{fig-model}(c, bottom). We say that after the $i$-th mutation (SNP) in one genome, the population 'skips' $\lambda_i$ previous variants. This implies that the set of segregating positions in each chromosome of a genome can be written as the series $\lambda_0 + \sum_{i} \lambda_i$, where $\lambda_0$ is the skip from the chromosome beginning, and $\lambda_i$ the subsequent. The case of constant lags is associated with a periodic function over genome positions, $seg(i) = seg(i+\lambda)$, where $seg$ is a binary or integral count variable indicating a segregation at site $i$. This predicts, in turn, periodic autocorrelation functions for each individual chromosome, with constant period $\lambda$ and frequency $\omega$ across population members. 

We start with the popular 1000 genomes dataset\cite{Auton:2015vv} (D.1). Fig.\ref{fig-1}(a) shows autocorrelation among all SNP positions, $seg(i)$, in the first and largest human chromosome (full left-arm, ${\sim}0.25$ billion SNPs for each individual), across all members of the European (EUR, $n=503$) and African populations (AFR, $n=661$). The left-to-right $y$-values indicate correlations at increasing genomic distances. The gray band shows autocorrelation confidence intervals (CI) among sites, and the dotted line shows the (min-max) range across all population members. These illustrate the strong regularity in correlations across sites, distances, and members. The solid horizontal line marks the baseline rate of $1/4$, indicating the proportion of sites that remain fixed at a time, as expected by NT.  

%, as would be expected from the common assumption of homogeneous mutation rates

%*** These plots illustrate how distinct mutation rates impart distinct correlational landscapes (structures)  across population members' position-to-position genomes, and, in turn, the number of pairwise differences in them. We will demonstrate that these increases in correlations are associated with patterned increases in effect observations for the same variants, $a,b,c...$, at increasing genetic backgrounds, $x_0$, and, consequently, with variants increased robustness. 

%simple calculation
%We thus expect their instantaneous frequency $\omega_{1}$ (i.e., of a single square column) to be $\omega_{1} \approx 6,200/247 = 25.1$ for chr-1. 

%This frequency-period combination suggests the diagonal 'hopping' pattern of (2), Fig.\ref{fig-model}(c). 
%This difference in patterns of selection is used as the first source of evidence for the LSM, followed by experimental interventions on selection in the next section. 

Fig.\ref{fig-1}(a) also indicates empirical frequencies (i.e., number of cycles per genome), $\hat{\omega}$, and lags, $\hat{\lambda}$, of mutations from all genomes in these populations. The EUR population has gone through the out-of-Africa bottleneck\cite{Henn:2012aa,Amos:2010uw} and, unlike the AFR, already underwent the second demographic transition. Multiple genome scans have identified in the past decades differential patterns of selection among these two populations\cite{Carlson:2005tl,Hawks:2007vo,Kayser:2003wa,Sabeti:2007ud,Courtiol:2013ty}, and considered the role of novel and diverse environments in this transition\cite{Akey:2004vr,Hawks:2007vo,Courtiol:2013ty}. Under higher selection pressure (e.g., higher windowed Tajima $D$ values\cite{Carlson:2005tl}), the observed number of cycles decreases, while observed lag periods ${\lambda}$ remain constant, Fig.\ref{fig-1}(a). The pattern can be seen in the populations of low-coverage sequencing datasets (D.1) and, even more clearly, in the subpopulations of high-depth datasets. Fig.\ref{fig-1}(e, bottom-row) shows number of cycles and periods in the African and British-born populations in the UK Biobank\cite{Halldorsson:2022ux} (D.2, $1000$ bootstraps, $6$ different sample sizes in $[500, 2000]$) for all chromosomes (and the associated significant results for same $\lambda$ and different $\omega$ hypothesis tests\cite{sm-anonymous}).

\subsection{Periodic Correlation.}

%was depicted in Fig.\ref{fig-model}(c), and 
Increased frequency $\omega$, with little change in period $\lambda$, leads to periodic correlation across genomes. The notion of correlated periodicity is well-studied\cite{Ghosh:1999vj}, outside genetics. It can be demonstrated in different ways. The colored strips in Fig.\ref{fig-1}(a, bottom) show population members (each a distinct color) ranked by correlation in each genome position. In periodically correlated systems, variants assume specific correlation levels periodically, one at a time. The strip depicts these across-genome patterns. Fig.\ref{fig-1}(c) shows graphical tests for correlation periodicity\cite{Hurd:1991ue} (D.1). In populations that are periodically correlated, observed frequencies in the test are distributed in equally-spaced diagonals. We observe this (in $\hat{\lambda}$ genome segments) for the AFR, but not EUR, population. 

According to the previous, systems $(2,3)$ differ in their frequency spectra. System $(2)$ is composed of exponential frequencies, and $(3)$ of exponential and fibonaccian frequencies. Fig.\ref{fig-1}(b) shows cumulative autocorrelations in $seg(i)-seg(i+l)$ vectors, where $i$ are SNP positions\cite{sm-anonymous}, and, $l$ are all integral lags $l \in [0,\hat{\lambda}]$. This is a common way of revealing periodicities in time-series. The EUR population has regular frequencies at $\sfrac{1}{\omega}$ intervals every $m$ variants. The AFR population has a further periodicity: a bisection every ${\sim}0.81{\times}m$ periods. Both are predicted by the LSM, Eq.(\ref{eq-posrates}). Fig.\ref{fig-1}(e, top-rows) show the same results in the UK Biobank (EUR, AFR). These results further suggest that systems $(2,3)$ are associated with populations, respectively, under selection, $(2)$, and drift, $(3)$. Distinct rates lead also to distinct autocorrelation patterns across systems, resembling peaks $(2)$ or triangles $(3)$. Practitioners will recognize these patterns, alternating present/absent triangles of same base length, from pairwise LD-block plots\cite{Todesco:2020vl,Gabriel:2002wn}. The patterns in Fig.\ref{fig-1}(b) do not correspond to those patterns, but to their generative source. For example, it is consistently observed that LD-blocks extends over distinct distances in AFR and EUR populations\cite{Frisse:2001uf,Reich:2001us}, and that these differences are associated with distinct statistical opportunities (e.g., detect associations in EUR, then perform fine-mapping in AFR\cite{Reich:2001us,Gabriel:2002wn}).

\begin{figure}
\centering
\includegraphics[width=1\linewidth]{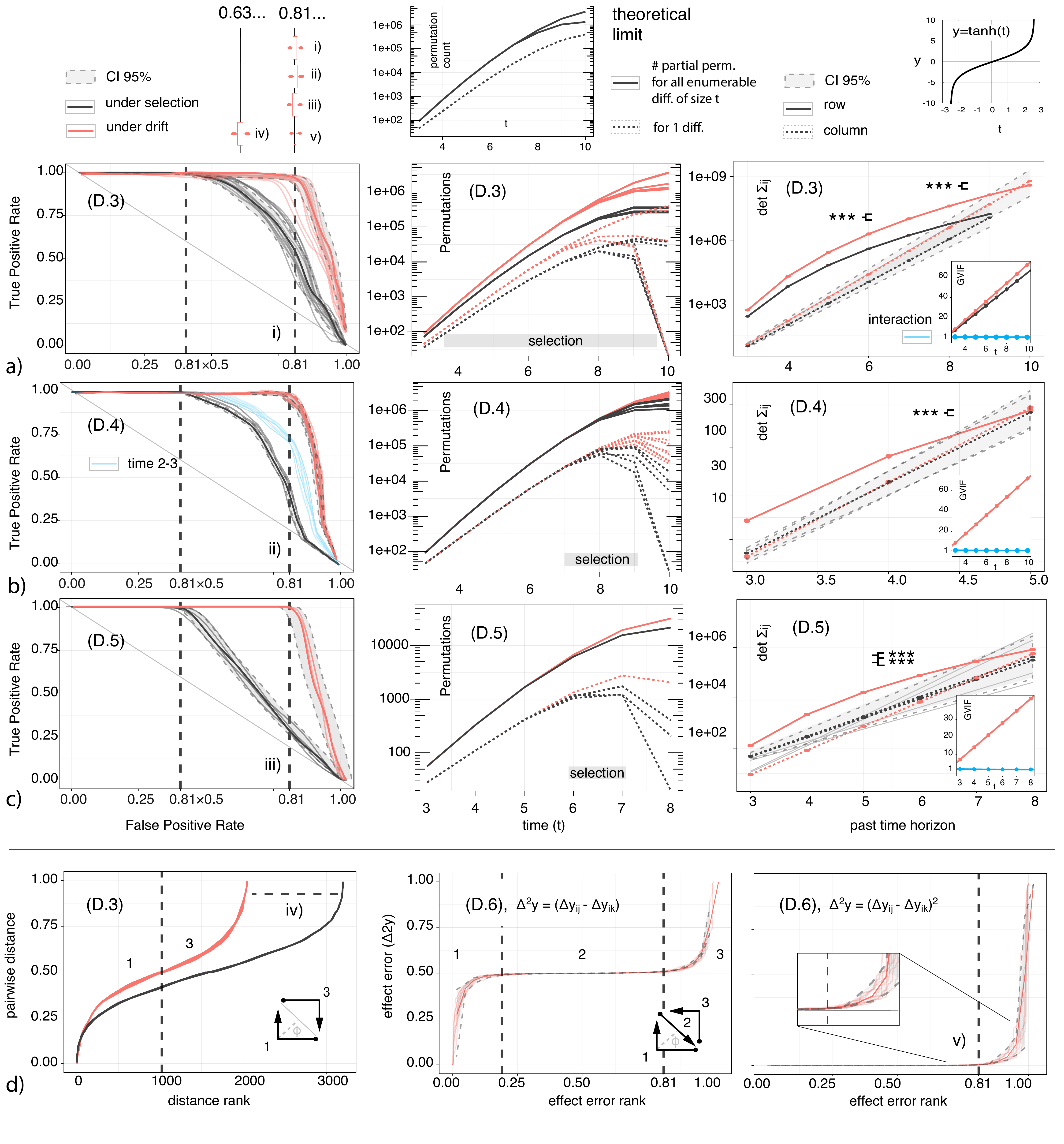}\\
{\small \refstepcounter{figure}\label{fig-2}\setstretch{0.5}\sffamily\noindent\textbf{Figure \arabic{figure}}\hspace{1em}\textbf{Experimental interventions on selection and effects (fine time-scale).}  Experiments over \textbf{(a)} yeast\cite{Wang:2018uf} (D.3), \textbf{(b)} fly\cite{Nosil:2020tf} (D.4) and \textbf{(c)} E.coli\cite{Nosil:2018vb} (D.5) populations, D.3 and D.5 have populations placed experimentally under selection and D.4 has high selection levels in a single interval, \textbf{(right)} Empirical ROC curves, where dashed lines and box-plot variation (i-iii, top-right panel) show LSM limits, Eq.(\ref{eq-posrates}), \textbf{(middle)}  theoretical (top panel) and empirical number of (partial) permutations for one and all enumerable differences of size $t$ (y-axis in log-scale), \textbf{(left)}  Generalized variance ($\det \Sigma_{ij}$) and their inflations (GVIF, lower panels), \textbf{(d)} \textbf{(left)} spatial distances among all pairs of genes in D.3 follow a $\tanh$ function (top-right panel) and $e$-sized drift-selection increases (boxplot, iv, top-right panel) as expected by the LSM, \textbf{(middle-right)} experimental effect observations in bi-mutation experiments\cite{Costanzo:2016vk} (D.6) follow LSM functionals and limits.}
\end{figure}

\subsection{Mutation Survival and Prediction.}

In this section, we consider how the previous rates limit the generalizability of mutation effects. Consider two populations with the same founder genome, one under drift and another under selection, evolving during a $[0,t_{max}]$ period. These populations will span distinct combinatorial and statistical landscapes for the selection process. Let $surv_t(x_0+a)$ be a binary indicator of a SNP $a$'s survival $t$ generations after its founding mutant. ROC curves are the most common analytical tools to understand the out-of-sample generalizability of algorithms and biological markers, plotting their true-positive (TP) vs. false-positive (FP) rates. The area-under-the-curve for the time-dependent ROC SNP survival curve indicates, in this case,

\begin{equation}\label{eq-surv}
Pr\, \big[\, surv_t(x_0+a)>surv_t(x_0) \; \big| \; y(x_0+a) > y(x_0), \,t\,\big],
\end{equation}

%,Hanley:1982te
% \cite{Blanche:2013uy,Beyene:2020tw} 
 and thus selection's ability to rank the true fitness gains in populations after $t$ experimental generations\cite{sm-anonymous} (i.e., with high TP rates). An elbow-curve in the ROC diagram, with maximal area and constant TP, describes Darwinian selection (where survival is a reliable indicator of fitness $y$ across mutations). The ROC relationship to natural selection is further discussed in \cite{sm-anonymous}. Fig.\ref{fig-2}(a-c, left-column) shows empirical ROC curves for variants in Saccharomyces cerevisiae\cite{Nguyen-Ba:2019vb} (D.3, artificial selection, 10 epochs of 100 generations, 2 population replicates), Drosophila melanogaster\cite{Rudman:wd} (D.4, observed high selection in 1 interval, 5 monthly intervals, 10 replicates), and Escherichia coli\cite{Zheng:2020uy} (D.5, artificial selection, 5 generations, 10 replicates) populations. Fig.\ref{fig-1}(d) shows, in turn, ROC curves with the previous observational datasets, D.1 and D.2, under random sampling. 

 Under uniform and noiseless selection mechanisms, true positive (TP) rates (y-axis) are associated with effects that are consistent across entire populations (and all possible $x_0$), Eq.(\ref{eq-surv}). Maintenance of 1.0 TP rates across growing populations in D.3-5 are thus indicative of effect invariance\cite{sm-anonymous}. In long time-scales, Fig.\ref{fig-1}(d), effects are invariant under selection until $1/4$ of the full sample, which coincide with their level of pairwise fixations. The increased genetic diversity rates under drift increases the number of background conditions in which effects are observed and, consequently, effect invariance for a larger section of the (combinatorially possible) populations. The rate where this transition happens is ${\sim}0.81\times m$, which is the rate necessary, Eq.(\ref{eq-posrates}), for variants to become simultaneously balanced (other, purely statistical cases, are discussed in \cite{ribeiro-ev}). This is seen across all data, D.1-5. By maintaining balance among mutations, increased genetic diversity rates allow effects to remain applicable throughout the diversifying populations, and their full evolutionary trajectories. This is true not only for long time-scale data, but also in fine time-scale experiments.

\subsection{Combinatorial Enumeration.}

% (a difference of size $1$ having $2$, and of $m$ having $m!$)

We now take a purely combinatorial perspective on adaptation, in contrast to the statistical of other results. According to the previous, we can characterize populations by the number of effect observations they generate or enumerate\cite{ribeiro-ev}. A homogeneous population generates none. Systems $(2,3)$ generate the maximum number of backgrounds under which a (limited) set of variant effects are observed. What we generally consider a genetic difference (e.g., in NT) is a fixed set of $t$ variants that change across a pair of population members. In a partial permutation, while this difference remains fixed, all other $m-t$ factors vary completely. The maximum number of backgrounds for a single difference, $(2)$, of size $t$ thus corresponds to all permutations of $m$ with $t$ positions fixed (i.e., all partial permutations for a single difference). The maximum number of backgrounds for all differences, $(3)$, corresponds to all permutations of $2^m$ unique differences (i.e., $\max [\pi] = \sum_t^m \binom{m}{t} = 2^m$). The latter is the number of enumerable differences, according to NT, for $m$ sites under drift. Fig.\ref{fig-2}(a-c, middle-column) show the number of enumerable permutations across time in D.3-5 (y-axis in log-scale). Curves for populations under drift and selection coincide with two combinatorial limits. On the top-left panel is the theoretical limit of permutations (solid line) for differences of size $t$. Experiment D.3 sustained a selective intervention (stressful, acidic environment) across its 10 time points, while D.4-5 had the intervention concentrated in one point (a natural seasonal change in D.4 and fluorescence-based selection in D.5). These periods are indicated with gray bars in Fig.\ref{fig-2}(a-c, middle-column).  In all cases, the empirical number of permutations had order-of-magnitude increases in the number of partial permutations - this difference sustained throughout the entire experiment in D.3. In fact, systems' empirical number of permutations under drift largely follow the theoretical number of permutations for all differences, Fig.\ref{fig-2}(a-c, middle-column) - a requirement for type $(3)$ systems.  With constant selection (D.3), the number of permutations per each difference under drift coincides with the number of permutations under selection at the end of the experiment. This indicates that the system can observe all permutations, and effect observations, per genetic difference and variant. Experiments with a single intervention, D.4-5, observe a steep increase in the number of partial permutations per difference at the time of intervention ($t=7$ and $t=6$) - many replicates going from the limit for a single difference to all differences in this single period. Fewer partial permutations, and steep decreases in the number of partial permutations, are observed in intervals after or under selection. This suggest that evolution is constantly pushing populations against (and is limited by) a combinatorial limit on effect observations and their backgrounds in both cases - with drift associated with limits for more concurrent mutations and genetic differences.  

\subsection{Generalized Variance.}

%This can be indicated by a null Generalized Variance (GV)\cite{sm-anonymous}, the determinant of the population's variance-covariance matrix, but only in the previous systems (i.e, with their rates and stationarity assumptions).  

The time-extended behavior of systems $(2,3)$ with rates in Eq.(\ref{eq-priorrates}) can be visualized with a Latin-Square matrix. Let the first matrix row correspond to a set $\{a,b,c,...,[m]\}$ of SNPs at an instant, and columns to environment-population time lags. A population change rate of $(m-\nicefrac{1}{m})$, increases the environment-population lag in one unit of time for all SNPs (thus, as expected, allowing a SNP $a$ to be evaluated against all environmental variations iteratively). This process repeats every $m$ steps, leading to the periodic matrix in Fig.\ref{fig-model}(d). It was discussed in detail in \cite{ribeiro-ev}, and is further reviewed in \cite{sm-anonymous}. In $(2)$ each environmental background for a SNP corresponds to a cell in the matrix's diagonal, and in $(3)$ to all diagonals (one for each SNP). The Generalized Variance (GV)\cite{Sen-Gupta:2006uk,sm-anonymous} is the determinant of a population's variance-covariance matrix. Since we defined effect invariance as low variance after all background variation, the GV of the previous matrices, each cell containing an effect observation, can be used as measure of effect invariance.  This follows simply from Leibniz permutation-based definition for determinants (i.e., the one taught at the high-school level algebra classes), and the previous definition of effect invariance.

%This is true only for the previous systems (i.e, with their rates and stationarity assumptions). 

%The Generalized Variance (GV) can describe effect invariance in systems (2-3)\cite{sm-anonymous}. 

Fig.\ref{fig-2}(a-c, right-column) show  sample variance-covariance determinants calculated independently for populations (row, solid line), $\det \Sigma_{row}$, and time (column, dotted line), $\det \Sigma_{col}$, across experiments. The figure shows that, as in all complete designs, row and column-wise GVs coincide\cite{Fox:1992vl}, $\det \Sigma_{row}$ = $\det \Sigma_{col}$. It implies, in turn, that across-diagonal variances between environment and populations coincide in $t_{max} \times t_{max}$ matrices (rightmost dots) - a requirement\cite{sm-anonymous}  for systems $(2,3)$. The figure also shows what happens as we reduce the time horizon (x-axis) progressively from $t_{max}$, from all $t_{max}$ generations until the founder to only $3$. This breaks the previous relations, demonstrating how the previous rates generates the populations structures depicted in Fig.\ref{fig-model}(d) across time, in both evolutionary regimes.

%adaptation can be seen as {time-extended} $m{\times}m$ designs, like the LSM, across both evolutionary regimes.

%Periodic lags $\lambda$ of size $m$ fulfill this condition for one or $m$ variants, Eq.(\ref{eq-posrates}).
The LSM corresponds to a fully-nested model for effect observations, where we rely not only on main effect estimates, but all interaction effects (thus an ANOVA\cite{ribeiro-ev} with all combinatorial interactions). In this interpretation, all enumerable interactions correspond to all enumerable effect observation backgrounds. Unbiasedness of effect estimation is therefore associated with the unbiasedness of interaction effects in typical designs. Only by keeping balance among background conditions, we can estimate robustness accurately. As noted by Fox and Monette\cite{Fox:1992vl} (Sect.6), while interaction effects are often ignored in practice, the GV inflation factor (GVIF) is a uniquely suited index to indicate the extent to which population imbalance will compromise interaction effect estimation. In balanced designs, we expect the GVIF of interactions to remain unitary, despite the otherwise large increases in colinearity. This is also shown in Fig.\ref{fig-2}(a-c, right-column, lower-panels), with interaction GVIFs shown in blue.    
%(and, as expected, linear)

\subsection{Knockout Effect Observations.}

The previous results showed the consequences of experimental manipulations on evolutionary regimes (same genome, different regimes). As a representational theory of effects, the LSM can also be evaluated with experiments that, instead, manipulate individual variants (and can thus measure, experimentally, their effects on fitness). Before those results, Fig.\ref{fig-2}(d, left) shows spatial distances among all pairs of genes in D.3 (normalized, in the same chromosome and experimental times). With the LSM assumptions, distance ranks should follow a $\tanh$ function\cite{sm-anonymous}, Fig.\ref{fig-2}(top-right). Distances among genes are $(1-\sfrac{1}{e}) = 0.63...$ times larger under selection, which corresponds to the $(1-\sfrac{1}{m})^m$ increase in rates illustrated in Fig.\ref{fig-model}(b,c) and Eq.(\ref{eq-posrates}).  The figure also shows this scaling factor's variation (iv, box-plot) across all gene pairs and times. 

%This is because the number of pairwise distances in a square follow the arithmetic series illustrated in Fig.\ref{fig-model}(c).

%This further supports the hypothesis that genes are organized in square-like distributions across populations, Fig.\ref{fig-1}.

%(observed at different times, and backgrounds, across the population)
%Gene knockout methods, like 
%CRISPR and 
Temperature-Sensitive (TS) gene knockout experiments intervene on a single gene, adding an allele $a$, to a genome $x_0$. The observed difference in fitness corresponds to a single effect observation, $\Delta y_{ij} = y(x_0+a) - y(x_0)$. Costanzo et al.\cite{Costanzo:2016vk} (D.6) measured such effects for all gene pairs $(ij)$ in yeast. Fig.\ref{fig-2}(d, middle) shows all pairwise differences among effects (normalized, same chromosomes). The plot (of effect differences) follows the same shape as the one for spatial differences, Fig.\ref{fig-2}(d, left), with an added plateau at $0.5$. The plateau corresponds to same, or near, position alleles that are not present in the single and fine experimental time of Fig.\ref{fig-2}(d, left). It also illustrates effect invariance, and the condition $(\Delta y_{ij} - \Delta y_{ik})^2=0$ across all effect differences from a fixed gene $i$\footnote{where $(jk)$ are all other same-chromosome genes.}. The plateau is also  reminiscent of FGM's frontier, where many effects take antagonistic values randomly.

%these correspond to squares'  lower-triangle, diagonal, and upper-triangle, Fig.\ref{fig-2}(d, lower-right diagrams). Each diagonal
We can divide the observation of effects for a given variant in 3 phases (before, after and in the plateau). According to the LSM population-wide representation, the rates in Eq.(\ref{eq-posrates}) represents 'costs' for increased robustness, or, more precisely, for effect observations under increased genetic backgrounds. Individual effect observations (left) lead to before-after only phases, and across-population observations (middle) to before-plateau-after. This is further illustrated in Fig.\ref{fig-2}(d, right) which shows \textit{squared} effect differences, and leads to the same $0.8{-}0.2$ limits seen previously, D.1-5. The figure also shows these distinct phases in the previous time-extended matrix representation, Fig.\ref{fig-2}(d, lower-right diagrams). We can now return to the FGM, whose chief argument was that evolution proceeds by small effect mutations. The argument was also its main source of criticism, starting from Kimura\cite{Kimura:1983tf}, who argued, theoretically, for the importance of intermediate effect sizes to adaptation. Contributions from Gillespie, Orr and Gerrish suggested, now including empirical components, that adaptation is characterized by a combination of exponential and small-sized 'jumps' in fitness ranks\cite{Gerrish:2001tl,Orr:2002ve,Rozen:2002uu}. The previous not only confirms these findings, but also gives them a specific $0.8{-}0.2$ effect size distribution. Additionally, the work here suggests how effects are spatially distributed across the genome, and how their sizes and spatial distributions change under selection.

\section{Conclusion}

%In conclusion,  patterns of sync and phase between populations and their environments can change the number of genetic backgrounds in which effects on members are observed, and levels of robustness across these populations. Rates in  Eq.(\ref{eq-posrates}) indicate the number of backgrounds that are observed for each SNP in systems, (2) with a single ($1$)  variant and no variation of other ($m-1$) variants, and (3) with all ($m$) variants and full variation of other ($m-1$) variants, in the same time interval. 

% under change
We formulated the hypothesis that adaptation processes systematically amplify the robustness of their mutations. Robustness is highest under specific patterns of lag and synchronicity between systems and their environments. We demonstrated these patterns in the broad time-scale of high-depth genomic datasets, and the fine-scale of multiple barcoding longitudinal experiments (in 6 different ways). Rates in  Eq.(\ref{eq-posrates}) indicate the number of backgrounds that are observed for each SNP in systems with $(2)$ no variation of other ($m-1$) variants, and, $(3)$ full variation of other ($m-1$) variants, in the same time interval. 
 
%[[proposes identifies set new and strict generalizability limits for general genetics and evolutionary research.]] 
 %(and their billions of standing variants)
The perspective sheds light on fundamental aspects of adaptation, such as its pace, limits and predictability. These limits have  implications across the full range of genetics and evolutionary research. The perspective allowed us to present a complete picture of how sets of whole-genomes change in response to environmental changes, and the essential role of standing variation and population structure in adaptation. The picture is one of adaptation as recursive  back-and-forths  between environments and populations, where rapidly changing environments prompt, in return, the need for systems to generalize gains across these new conditions.  We believe this theoretic-empirical perspective could help transform our understanding of evolution, biodiversity maintenance and medical human genomics. 

\section*{References}
\bibliography{bib}

\end{document}